\documentclass[runningheads,a4paper]{llncs}

\usepackage{etex}
\usepackage{wrapfig}

\usepackage{graphicx}
\graphicspath{{figures/}{figures//}}
\usepackage[colorinlistoftodos,prependcaption,textsize=tiny]{todonotes}
\usepackage{lipsum} 

\usepackage{cite}
\usepackage{makecell}

\usepackage{subfigure}

\usepackage{amsmath}
\usepackage{amssymb}

\usepackage{tabularx}
\usepackage{color}
\usepackage[hyphens]{url}
\usepackage[noend]{algorithmic}
\usepackage{paralist}
\usepackage[all]{xy}

\usepackage{xspace}
\newcommand{\etal}{\textit{et al.}\xspace}
\newcommand{\etc}{\textit{etc.}\xspace}
\newcommand{\ie}{\textit{i.e.,}\xspace}
\newcommand{\eg}{\textit{e.g.,}\xspace}

\newcommand{\aka}{\textit{a.k.a.}\xspace}
\mathchardef\-="2D

\newcommand{\framework}{\textsl{\mbox{Oblivion}}\xspace}

\renewcommand{\paragraph}[1]{\medskip \noindent \textbf{#1.\ }}

\newcommand{\vk}{\ensuremath{\mathsf{vk}}}

\newcommand{\sk}{\ensuremath{\mathsf{sk}}}
\newcommand{\accept}{\ensuremath{\mathsf{accept}}}
\newcommand{\reject}{\ensuremath{\mathsf{reject}}}

\newcommand{\ot}{\ensuremath{\overset{\$}{\leftarrow}}\xspace}

\begin{document}

\title{Oblivion: Mitigating Privacy Leaks by Controlling the Discoverability of Online Information}
\titlerunning{Oblivion: A Framework for Enforcing the Right to be Forgotten}

\author{Milivoj Simeonovski\inst{1} \and Fabian Bendun\inst{1} \and Muhammad~Rizwan~Asghar\inst{2}\thanks{This work was done when the author was at CISPA, Saarland University, Germany.} \and \\ Michael Backes\inst{1,3} \and Ninja Marnau\inst{1} \and Peter Druschel\inst{3}}
\authorrunning{Simeonovski, Bendun, Asghar, Backes, Marnau and Druschel} 
\tocauthor{Milivoj Simeonovski, Fabian Bendun, Muhammad~Rizwan~Asghar, Michael Backes, Ninja Marnau, Peter Druschel}
\institute{CISPA, Saarland University, Germany\\
\and The University of Auckland, New Zealand\\
\and Max Planck Institute for Software Systems (MPI-SWS), Germany\\
}

\maketitle

\begin{abstract}

Search engines are the prevalently used tools to collect information about individuals on the Internet. Search results typically comprise a variety of sources that contain personal information --- either intentionally released by the person herself, or unintentionally leaked or published by third parties without being noticed, often with detrimental effects on the individual's privacy. To grant individuals the ability to regain control over their disseminated personal information, the European Court of Justice recently ruled that EU citizens have a \emph{right to be forgotten} in the sense that indexing systems, such as Google, must offer them technical means to request removal of links from search results that point to sources violating their data protection rights. As of now, these technical means consist of a web form that requires a user to manually identify all relevant links herself upfront and to insert them into the web form, followed by a manual evaluation by employees of the indexing system to assess if the request to remove those links is eligible and lawful.

In this work, we propose a universal framework \framework to support
the automation of the \emph{right to be forgotten} in a scalable,
provable and privacy-preserving manner. First, \framework enables a
user to automatically find and tag her disseminated personal
information using natural language processing (NLP) and image recognition techniques and
file a request in a privacy-preserving manner. Second, \framework
provides indexing systems with an automated and provable eligibility
mechanism, asserting that the author of a request is indeed affected
by an online resource. The automated eligibility proof ensures censorship-resistance so that only legitimately affected
individuals can request the removal of corresponding links from
search results. We have conducted comprehensive evaluations of \framework, showing that the framework is capable of handling 278 removal requests per second on a standard notebook (2.5 GHz dual core), and is hence suitable for large-scale deployment.

\end{abstract}




\section{Introduction}

The Internet has undergone dramatic changes in the last two decades,
evolving from a mere communication network to a global multimedia
platform in which billions of users not only actively exchange
information, but also increasingly carry out their daily personal
activities. While this transformation has brought tremendous benefits
to society, it has also created new threats to online privacy that
existing technology is failing to keep pace with. In fact, protecting
privacy on the Internet remains a widely unsolved challenge for users,
providers, and legislators alike. Users tend to reveal personal
information without considering the widespread, easy accessibility,
potential linkage and permanent nature of online data. Many cases
reported in the press indicate the resulting risks, which range from
public embarrassment and loss of prospective opportunities (\eg when
applying for jobs or insurance), to personal safety and property risks
(\eg when stalkers, sexual offenders or burglars learn users'
whereabouts).

Legislators have responded by tightening privacy regulations. The
European Court of Justice recently ruled in Google Spain v. Mario
Costeja Gonz\'{a}lez~\cite{EuCourtJudgment} that EU citizens have a
fundamental \emph{right to be forgotten} for digital content on the
Internet, in the sense that indexing systems such as Google (or other
search engines, as well as systems that make data easily discoverable,
such as Facebook and Twitter) must offer users technical means to
request removal of links in search results that point to sources
containing their personal information and violating their data
protection rights.\footnote{In the court's case, the plaintiff
  requested the removal of the link to a 12-year old news article that
  listed his real-estate auction connected with social security debts
  from the Google search results about him. The court ruled that the
  indexing by a search engine of the plaintiff's personal data is
  ``prejudicial to him and his fundamental rights to the protection of
  those data and to privacy --- which encompass the \emph{right to be
    forgotten} --- [and overrides] the legitimate interests of the
  operator of the search engine and the general interest in freedom of
  information.''} While a comprehensive expiration mechanism for
digital data has often been postulated by privacy advocates in the past, this
court decision, for the first time, imposes a legal constraint for
indexing systems that operate in the EU to develop and deploy suitable
enforcement techniques. As of now, the solution deployed by leading
search engines, such as Google, Microsoft and Yahoo, consists of a
simple web form that requires a user to manually identify all relevant
links herself upfront and to insert them into the web form, followed
by a manual evaluation by the search engine's employees to assess
whether the author of the request is eligible and the request itself
is lawful, \ie the data subject's right to privacy overrides the
interests of the indexing operator and the freedom of speech and
information. 

According to the Google transparency report~\cite{Google:2015:Removals}, the number of removal requests that have been submitted to Google since the court
decision in May 2014 has already exceeded 1/5 of a million and the
number of URLs that Google has evaluated for removal are approximately
3/4 of a million. Clearly, in order to
enable efficient enforcement, it is essential to develop techniques
that at least partly automate this process and are scalable to
Internet size, while being censorship-resistant by ensuring that
malicious users cannot effectively blacklist links to Internet sources
that do not affect them.

\paragraph{Our Contribution}
We propose a universal framework, called \framework, providing the
foundation to support the enforcement of the \emph{right to be
  forgotten} in a scalable and automated manner. Technically,
\framework provides means for a user to prove her
eligibility\footnote{With our framework we allow for the automation of
  the eligibility proof of the user. Eligibility in our framework
  describes the user being personally affected by an online source, or
  in legal terms being the \emph{data subject}. The \emph{right to be
    forgotten} additionally requires that the user's data protection
  rights override the legitimate interests of the search engine
  operator and the freedom of information. This assessment of the
  \emph{lawfulness} of the request is a purely legal task, which is in
  the domain of courts. Hence the technical assessment of lawfulness
  is out of scope for our framework. If courts and regulators agree on
  guidelines for this assessment, \framework could be extended to a
  partly automated assessment of these guidelines in future work.} to
request the removal of a link from search results based on trusted
third party-issued digital credentials, such as her passport or
electronic ID card.

\framework then leverages the trust imposed by these credentials to generate eligible removal requests. More specifically, the officially-generated signatures contained in such credentials comprise personally-identifiable information of the card owner, such as her signed passport picture, address, \etc These so-called signed \emph{attributes} are subsequently automatically compared with publicly available data whose removal should be requested, in order to determine if a source indeed contains information about a given entity. In \framework, we use state-of-the-art natural language processing (NLP) and image recognition techniques, in order to cover textual and visually identifiable information about a user, respectively. Further modalities can be seamlessly integrated into \framework. These techniques in particular automate the task for a user to determine if she is actually affected by an online source in the first place. The outcome of these comparisons, based on the signed attributes, is then used to provide proof to the indexing system that a user is eligibly affected by a source. To avoid creating further privacy concerns, \framework lets the user prove her eligibility to request data removal without disclosing any further personal information beyond what is already available at the link. This approach applies to a variety of different indexing systems, and in particular goes beyond the concept of search engines that we refer to throughout the paper for reasons of concreteness. Moreover, \framework exploits the homomorphic properties of RSA~\cite{RSA} in order to verify the eligibility of an arbitrarily large set of user credentials using only a single exponentiation, and is thus capable of handling 278 requests per second on a standard notebook (2.5 GHz dual core and 8 GB RAM). We consider this suitable for large-scale~deployment.

\paragraph{Outline}
This paper is structured as follows. We review related work in Section~\ref{sec:relatedwork}. The conceptual overview of \framework and its detailed realization are presented in Sections~\ref{sec:framework} and~\ref{sec:details}, respectively. Section~\ref{sec:performance} provides performance analysis of \framework. Section~\ref{sec:discussion} discusses various aspects of \framework. Next, we conclude and outline future work in Section~\ref{sec:conclusion}. Appendix~\ref{sec:security} formally states and proves the security properties of \framework.


\section{Related work}
\label{sec:relatedwork}
The most common way to prevent \emph{web robots} (or \emph{web crawlers})~\cite{OlNa10_FoundinIR} from indexing web content is \emph{the Robots Exclusion Protocol} (\aka~robots.txt protocol)~\cite{robots.txt}, a standard for controlling how web pages are indexed. Basically, robots.txt is a simple text file that allows site owners to specify and define whether and how indexing services access their web sites. The use of this protocol for privacy enforcement is limited, since the file that defines the protocol can only be placed and modified by the administrator of the web site. The individual whose personal data is being published is hardly capable of contacting and persuading all administrators of these sources to remove the data or modify the robots.txt file. 
There are many attempts to approach this privacy enforcement problem in an orthogonal fashion, by adding an expiration date to information at the time of its first dissemination~\cite{Xpire2, vanish, ReDuRevocation, ephemerizer, Castelluccia:2011:EphPub, Nair:2007:Ephemerizer}. The basic idea is to encrypt images and make the corresponding decryption key unavailable after a certain period of time. This requires the decryption key to be stored on a trusted server, which takes care of deleting the key after the expiration date has been reached. Although some of the approaches utilize CAPTCHAs to prevent crawling the images easily, there is no fundamental protection against archiving images and corresponding keys while they are still openly available, even though first successes using trusted hardware to mitigate this data duplication problem have been achieved \cite{Xpire2}. Another approach in this direction is the concept of sticky policies~\cite{MontSticky,PearsonStickey, Super-sticky, enforceSticky}. The concept was originally introduced by Mont~\etal \cite{MontSticky} and requires a machine-readable access policy to be bound to the data before it is disseminated. The policy then ensures that the recipient of the data acts in accordance with the policy definition. However, enforcement of such policies has to be backed by additional underlying hardware and software infrastructure. In addition to these shortcomings, a user needs to take care to augment data with expiration dates before the data is disseminated in all these approaches. Thus these approaches are inherently unsuited to cope with data that is already openly available on the Internet or gets published by third parties.
Finally, to implement the European Court of Justice's decision, Google, Microsoft and Yahoo recently launched dedicated web forms~\cite{GoogleRemoval, BingRemoval, YahooRemoval} for submitting removal requests. Users have to manually identify all relevant links and insert them into this form. Subsequently, the request is evaluated manually by the employees of the indexing system to assess first, weather the author is eligible to file that request and second, whether the link to the source needs to be deleted for a specific search. To this end, users have to additionally hand over a legible copy of an ID document. The necessity of handing over a user's full identity to use the service comes with additional privacy implications that one would like to avoid. \framework constitutes a technical follow-up to this solution, with a dedicated focus on censorship-resistance, while additionally avoiding the detrimental effect of having to disseminate further personal information.



\begin{figure*} [htp]
\centering
\includegraphics[width=.8\textwidth]{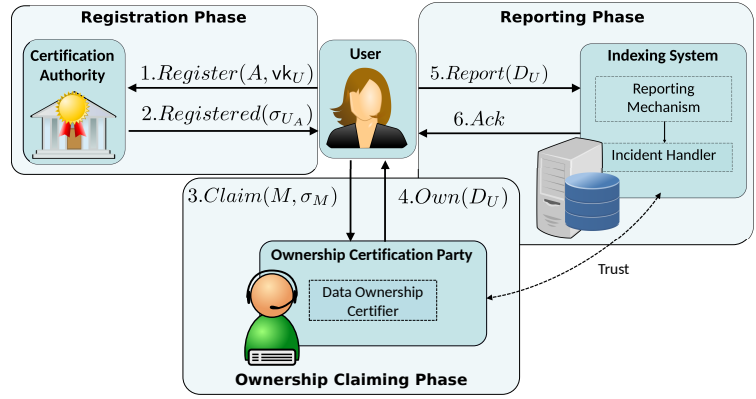}
\caption{Conceptual Overview of \framework.}
\vspace{-7mm}
\label{fig:general_framework}
\end{figure*}

\section{Conceptual Overview of Oblivion}

\label{sec:framework}
In this paper, we propose a framework laying the foundation for a
privacy-preserving automation of the \emph{right to be forgotten} in a
scalable manner. The basic idea is that users automatically identify
online sources that contain their personal data and can automatically
request its removal from indexing systems, if it violates their data
protection rights. Upon receiving the request, we enable the indexing
service to automatically verify if the author of the request is
provably affected by the source in question. Our framework is
sufficiently generic to incorporate any type of data, such as text,
pictures, voice and video. For brevity reasons, in this paper, we
mainly focus on two data types: text and pictures.

\subsection{Motivating Scenario and System Model}
\label{sec:scenario}
We start with a motivating scenario to explain the required
functionality of the framework and the different parties involved. We
assume that a user, Alice, discovers that an indexing service, say
Google, returns certain query requests with links pointing to a
document that contains her personal information and violates her
privacy. In the next step, Alice contacts an Ownership Certification
Party (OCP) in order to receive validation that this source indeed
contains her personal information. Such an OCP could be a third party
or the Google helpdesk. Along with the relevant links, she hands over
publicly verifiable ID documents such as driver's license, passport or
national ID card to the OCP. If the provided documents and the content
of the article in question indeed match (which will be automatically
checked by \framework), the OCP hands back a corresponding
certificate. Alice then contacts Google to request removal of these
links, providing an additional explanation, and proves her eligibility
to do so based on the certificate of the OCP. Upon receiving this
information, Google checks if the considered document is indeed
indexed by Google, and if the OCP certificate is valid for this
specific document and user. In this case, the requested article will
be removed from the indexing system.
 
Based on this use case scenario, we consider the following entities in
our proposed framework designed for automating the process of handling
removal requests.

\begin{description}

\item[User:] An authorized user who issues the request to
  remove her personal data.

\item[Indexing system:] This system is capable of removing links to sources
  containing a user's personal data from its indexing system, based on
  a removal request of the user.

\item[Ownership Certification Party (OCP):] It is responsible for
  verifying if the user is the eligible data subject of the source
  under consideration.\footnote{Ownership in this context should not
    be confused with the legal term. Legally, the OCP can only assess
    and certify the individual's eligibility since, at least in EU
    context, legal ownership is not applicable to the right to be
    forgotten.}

\item[Certification Authority (CA):] It issues publicly verifiable credentials to the users.

\end{description}

\subsection{Threat Model and Security Objectives}
\label{sec:goals}
We assume that all entities in the system fully trust the CA. However,
a CA does not need to be online because the issuance of credentials to
the users takes place out of band, typically for a longer period of
time, say a couple of years.

Unlike the CA, the OCP is an entity that is not fully trusted from the
user's perspective because it can try to learn the user's keying
material and additional user credentials not required for the
ownership verification; moreover, it might want to forge removal
requests.  The OCP is the only entity that is not part of the
traditional system. The OCP can be run by the organization (\eg
Google) that manages the indexing system, or it can be a third-party
service. The OCP is assumed to be online during the execution of a
request.

The indexing system is an entity inherently present in the traditional
system. The indexing system and the OCP mutually trust each other; in
practice, this is often trivially the case since the OCP and the
indexing system are often managed by the same organization. If the OCP
is an independent third party, this trust would typically be
established via the CA using appropriate certificates.

We assume that users protect their private keys or at least, if their
private keys are lost or stolen, a key revocation mechanism is
installed and the user generates new keys. During the ownership
verification, we do not assume any interaction between the users and
the OCP. A user can present the OCP-signed proof to remove links to
the data from multiple indexing systems, such as Google and Yahoo. We
also consider an external adversary that could harm credibility of the
user through replay attacks with the intention to make the service
unavailable. For providing confidentiality over the communication
network, we assume the presence of secure channels (such as
SSL/TLS~\cite{TLS}) between a user, the OCP and the indexing system.

Based on these assumptions, we intend to achieve the following
security objectives:

\noindent\begin{itemize}
\setlength{\itemsep}{1pt}
\setlength{\parskip}{0pt}
\setlength{\parsep}{0pt}
\item \emph{Minimal Disclosure:} An indexing system should not learn
  anything beyond what is required for eligibility checking and
  assessment of lawfulness. The court decision ruled that the right
  has to be judged on a case-by-case decision. Hereby, the right of
  the individual has to be balanced with the public right of
  information. Our system handles removal requests that prove eligibility but do not reveal any further information beyond what
  can be found in the online source in question.\footnote{Although
    \framework provides for minimal disclosure, the indexing system
    might request additional information, such as an author's name,
    for liability reasons in a real-world deployment of
    \framework. Moreover, the assessment of lawfulness could in some
    cases also require additional personal information.}

\item \emph{Request Unforgeability:} The system should be designed
  such that an indexing system can only verify user requests without
  any possibility of forging existing or generating new requests
  on behalf of the user.

\item \emph{Censorship-Resistance:} The system should prevent
  censorship in the sense that only requests from provably
  affected users should be taken into account.
  
\end{itemize}

In addition to ensuring these security properties, the system should
satisfy the following system properties in order to be suitable for
large-scale deployment. It should be \emph{scalable} in order to be
able to process a large amount of queries simultaneously, while at the
same time ensuring a thorough treatment of each individual query. It
should \emph{blend seamlessly into existing infrastructures}, to
enable adoption by current indexing systems and certification
authorities; moreover, the solution should be conceptually independent
of the device and the operating system used. Finally, it should be
\emph{easy to understand and use} even for the general public.

\subsection{Key Ideas of the Protocol}
\label{sec:approach}
\framework is built on top of already available infrastructure (as
explained in Section~\ref{sec:scenario}) that includes users, an indexing
system and a CA. For the automatic verification of ownership, we
introduce only a single new entity, the OCP, thus making our framework
deployable in practice. In the framework, we distinguish three main
phases: registration, ownership claim and reporting
phases. Figure~\ref{fig:general_framework} presents the overall
architecture for achieving the goals defined in
Section~\ref{sec:goals}.

\paragraph{Registration Phase}
During the registration phase, each user registers with the CA as
shown in Figure~\ref{fig:general_framework}. For the registration, a
user presents (in Step~1) her attributes (along with evidence) and
her verification key. The verification key should, for privacy reasons,
be generated by the user herself before contacting the CA, but the
generation of the key is orthogonal to our framework. The CA checks
the validity of the attributes presented, certifies them and returns
(in Step~2) a list of signed attributes, where each signed attribute
is bound with the user's verification key. Typical examples of
attributes are the date of birth, name or a user's profile picture.

\paragraph{Ownership Claim Phase}
Once a registered user finds leakage of her personal data through the
indexing system, she can contact the OCP claiming eligibility
(in Step~3). This is the core phase in which the OCP expects
justification of why the given piece of data affects the user. To make
such a justification, the user can put tags on the given data that
consist of her attributes which were signed by the CA. In order
to improve usability, we automate the tagging and
verification. One trivial automation method is to simply check if any
user attribute appears anywhere in the article; if this happens, the
matched item could be tagged with that attribute. The name attribute,
say Alice, could be matched in this way.

The exact matching can semi-automate the tagging process but it cannot
work in general because it may not return the correct results for all
user attributes. Let us consider a user attribute in the form of a
tuple: \textbf{$\langle$Nationality, German$\rangle$} (as explained in
Section \ref{sec:registration}). In order to match this attribute, the
OCP has to check if the user attribute or its synonym has appeared in
the article. This includes semantically linkable formulations, such as 
\emph{being a citizen of Germany} and \emph{having German nationality}.

Letting the user manually deploy this solution, \ie forcing the user
to find synonyms of each possible word in the article, is an
exhaustive task. Therefore, we employ an NLP-based technique --- the
named entity recognizer (NER)~\cite{NER} in our case --- for
efficiently collecting all possible candidates in the article. The NER
detects and classifies the data into various categories, such as
person, organization, location, date and time, and it thus helps to
identify if a user has attributes belonging to the category
identified by the NER. If yes, we can perform exact matching or run a
synonym checker~\cite{WordNet} on identified categories.
Articles containing a user's picture are tagged in a corresponding
manner.

After the attributes are matched, the user has to generate a proof by
preparing a message that contains a list of signed attributes that are
required for the verification, the tagged article and her verification
key. The user signs this message and sends it to the OCP (in Step~3)
as an eligibility claim.

The OCP first verifies the message signature and the signed attributes
used in the tagging.  If the claim relates to text attributes, the OCP
runs an entity disambiguator to identify whether the article is about
the user.  If the claim includes a picture, the OCP runs a
corresponding face recognition algorithm. Upon successful evaluations
of all steps, the OCP presents to the user an ownership token (in
Step~4).

\paragraph{Reporting Phase}
After receiving the ownership token from the OCP, the user sends a
request for removal to the indexing system (in Step~5). The indexing
system automatically validates the ownership token and then assesses
whether to remove the links pointing to the user's personal
information from its system. Finally, it sends (in Step~6) an
acknowledgment to the user, which could be a \emph{success} or
\emph{failure} message.



\section{Realization Details of Oblivion}
\label{sec:details}
In this section, we provide details of each phase of our framework and
explain the communication protocol to show interaction between
different components. An indexing system and a user are denoted with
IS and U, respectively. The communication protocol steps, described in
this section, correspond to the flow illustrated in
Figure~\ref{fig:general_framework}. After that, we provide details on
how to securely and efficiently realize the proposed protocols using
cryptographic primitives.

\subsection{Registration Phase}
\label{sec:registration}
As we can see in the communication protocol, a user sends (in Step~1)
her attributes, $A = \{a_1, a_2, \ldots, a_n \}$, which
characterize her, with supporting proofs and the verification key
$\vk_U$ to the CA. Each user attribute $a_i \in A$ is a name and
  value key pair \textsf{$\langle$NAME, VALUE$\rangle$}, representing
  name of the attribute and value specific to each user,
  respectively. For instance, an attribute name could be
  \emph{National}, and if say, a user is national of Germany, then the
  value will be \emph{German}. Some general user attribute names
  include, but are not limited to, \emph{Full Name}, \emph{Date of Birth},
  \emph{Place of Birth}, \emph{Current Residence} and \emph{ID
    Picture}.

Upon a successful verification of the provided data, the CA issues a
list of signed attributes $\sigma_{U_A} = \{ \sigma_{U_{a_1}},
\sigma_{U_{a_2}}, \ldots , \sigma_{U_{a_n}} \}$ and sends it back to
the user (in Step~2). Our attribute signing scheme binds every user's attribute with her verification key. Note that one
of the attributes $a_i$ is a profile picture that uniquely identifies
the user.

Steps~1 and 2 constitute the registration phase that takes place
securely and out of the band. The concept of digital signature
together with user attributes (signed by the government) is already
present in some EU countries~\cite{GermanID,EuID,EuSignatures}.

\subsection{Ownership Claim Phase}
In order to make an ownership claim to the OCP, we consider a user
client, say a browser plugin. The plugin sends the claim to the OCP
and receives an ownership token from the OCP in the case the claim can
be verified, cf. Figure~\ref{fig:general_framework}. In order to do
so, the first step is that the user client has to formulate the claim,
then it has to identify personal information and finally the actual
removal request has to be generated. In the next step, the OCP has to
verify the request. This is done by first verifying the authenticity
of the request and second verifying the relationship to the data. The
latter verification depends on the type of data, \eg face
recognition can be used for pictures. The last step is to generate the
ownership token that is then transferred from the OCP to the user. In
the following, we present the details of all these tasks.

\begin{figure} [htp]
\centering
\includegraphics[trim=85mm 0mm 0mm 100mm,clip,width=.6\textwidth]{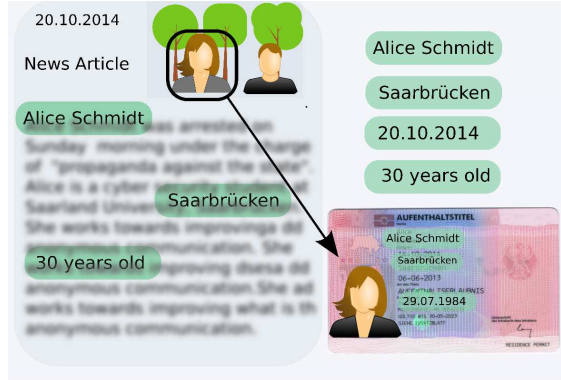}
\caption{An article illustrating personal information of Alice Schmidt who has an ID card with digital credentials issued by the German government.}
\label{fig:example}
\end{figure}

\paragraph{Identifying Personal Information}
For identifying user's personal information in an article (as
illustrated in Figure~\ref{fig:example}), a user client may run the
NER algorithm locally (assuming it is delivered as a part of the user
client) to extract all possible candidates. The NER algorithm could
also be run as a third-party service (\eg a web service), called by
the user client. After running the NER algorithm, a user client picks
each of the candidates and matches them with the user attributes (see
Figure~\ref{fig:example}). 

If the match is not successful, a user client runs a synonym
checker. If both words are synonyms then they are considered matched;
otherwise, the next candidate is picked from the queue for the
comparison. The synonym checker could be delivered as a part of the
user client. To make the user client lightweight, we can assume a
third party service (\eg a web service). In either case, the synonym
checker should be very specific to the attributes issued
by the CA.\footnote{
For instance, the user's date of birth might appear differently in an article, \ie in the form of her age as shown in Figure~\ref{fig:example}. If this happens, the age could be compared with the difference of the user's date of birth and publication date of the article, if present. As we can see in the example, \emph{30 years old} will be compared with \emph{20.10.2014 - 29.07.1984}. Further tests for checking syntactic equivalence are conceivable, but are postponed to future work.}

\paragraph{Face Detection}
Besides the textual description, an article could also contain a
user's picture, either as a solo or a group picture. Like textual attributes, the user client can run the face detection algorithm to automatically detect the user's face. On successful detection, a user client can automatically include the CA-signed user picture in the removal request, which is explained next.

\paragraph{Generating Removal Request}
After identifying personal information, a user client prepares a
removal request. During the preparation, it chooses all signed
attributes required for the ownership claim. Next, it packs them as
$P_{\sigma_{U_{A^*}}}$ so that the OCP can verify the signed user
attributes using a single exponentiation operation using the CA
verification key. This would also require a user client to include in
the message a subset of her attributes $A^*$ corresponding to the
packed ones, \ie $P_{\sigma_{U_{A^*}}}$. Since a user client signs the
message using the user's signing key, the user's verification key
$\vk_U$ is also included in the message to let the OCP verify the
message. For preventing replay attacks, a timestamp $TS$ is also
included in the message. The user client sends to the OCP (in Step~3)
the message $M=(TS, \vk_U, A^*, P_{\sigma_{U_{A^*}}}, D)$ along with
the signature $\sigma_M$.

\paragraph{Verifying Removal Request}
Upon receiving a removal request, an OCP verifies it before issuing
any ownership token. As a first step, the signature $\sigma_M$ over
the message $M$ is verified. Next, the OCP checks the timestamp and
verifies the packed version of the user attributes signed by the
CA. Then, the OCP checks if all tagged attributes are valid. This step
comprises the exact matching and/or synonym checking.

\paragraph{Face Recognition}
Optionally, the face recognition algorithm could be run provided there
is a user picture in the article. As we explained earlier in this
section, faces are pre-identified by the user client, in order to ease the job of the OCP. The OCP compares the user-tagged face with one provided as a signed user
attribute in the request (see Figure~\ref{fig:example}). If the face
recognition algorithm discovers similarity with a certain confidence,
the user's picture in the article is considered matched with her
profile picture.

\paragraph{Entity Disambiguation}
When the given article contains text, the OCP can execute the
disambiguation algorithm (\eg AIDA~\cite{AIDA}) for ensuring the
eligibility goal, \ie checking
whether the article is about the user. The outcome of this algorithm
is the relation between the user attributes, her name in particular,
and the context of the text. The outcome, say satisfying the
predefined threshold value, would help the OCP to mark the user as
being affected by the data in the article. Figure~\ref{fig:example} illustrates
an example article about Alice Schmidt.

\paragraph{Issuing Ownership Token}
On successful evaluations of all the steps performed by the OCP, the
user is issued an ownership token. This is accomplished by the OCP by
sending (in Step~4) an ownership token $D_U$ to the user. It is
important to note that the OCP verification protocol is
non-interactive.

\subsection{Reporting Phase}
Once the user receives the ownership token, she can report to the
indexing system. In this phase, a user reports by sending (in Step~5)
the ownership token $D_U$ (corresponding to $D$) to the indexing
system. The indexing system verifies the token, fires the incident and
sends (in Step~6) an acknowledgment $Ack$ to the user. If the OCP is
a third-party service, the ownership token is signed by the OCP and
could be sent to multiple indexing systems simultaneously.



\subsection{Efficient Cryptographic Realization} 
\label{sec:cryptoconstruct}
\label{sec:construction}

The cryptographic instantiation relies on RSA-full-domain hashing as
the underlying signature scheme. We briefly recall the definition of this
signature scheme. The scheme assumes a given collision-resistant
family of hash functions $\mathcal{H}_k: \{0,1\}^* \rightarrow
\{0,1\}^k$. In the following, we omit the security parameter $k$ for
readability.  The key generation \textsf{KeyGen} computes a key pair
$(\sk, \vk)$ by first computing an RSA modulus $N = p \cdot q$, where
$p$ and $q$ are two random primes, and then computing $e$ and $d$ such
that $e \cdot d = 1 \mod (p-1)(q-1)$. The keys are $\sk = (d, \vk)$
and $\vk=(e,N)$. The signing function \textsf{Sign(\sk, $M$)} computes
$\sigma_M := \mathcal{H}(M)^d \mod N$. Finally, the verification
function \textsf{Verify(\vk,$\sigma$,$M$)} outputs $\accept$ if
$\mathcal{H}(M) = \sigma^e \mod N$ and $\reject$ otherwise.

Using this cryptographic primitive, we finally describe the
construction that we propose for our framework to achieve goals
defined in Section~\ref{sec:goals}. The censorship-resistance and
eligibility checking goals could be achieved using X.509 based schemes
\cite{rfc3280}; however, those schemes are not able to achieve goals
including minimal disclosure (\ie disclosing only those attributes
required for the ownership claim) and scalability (\ie reducing
computational overhead on the OCP end). Using our construction, the
user can provide a minimal set of her attributes required for the
ownership claim, and we are able to delegate some computation to the
user client so that the OCP could be offloaded. Our construction
allows an OCP to verify all user attributes with just a single
exponentiation.

\begin{figure*}[ht!]
\setlength{\extrarowheight}{1em}
\begin{tabularx}{\textwidth}{|p{.492\textwidth}|p{.492\textwidth}|}
  \hline \textsf{Init($k$):} The system is initialized by running
  \textsf{RSA-Sig.Init($k$)}, which returns $\mathcal{H}(.)$.  &

  \textsf{KeyGen($1^\lambda$):} It runs
  \textsf{RSA-Sig.KeyGen($1^\lambda$)}. For the certification
  authority $\mathsf{CA}$, it returns $\langle \sk_\mathit{CA},
  \vk_\mathit{CA} \rangle$. For each user $U$, it returns $\langle
  \sk_U, \vk_U \rangle$.\\

  \textsf{CA.SignA($\sk_\mathit{CA}$, $\vk_U$, $A$):}
  Given the CA's
  signing key $\sk_\mathit{CA}$, the user's verification key $\vk_U$
  and a list of user attributes $A = \{a_1, a_2, \ldots, a_n \}$, the
  certification authority $\mathsf{CA}$ returns the list of signed
  attributes $\sigma_{U_A} = \{ \sigma_{U_{a_1}}, \sigma_{U_{a_2}},
  \ldots , \sigma_{U_{a_n}} \}$, which represents a list of signed
  attributes that belong to the user $U$. For each attribute $a_i \in
  A$, it computes $\sigma_{U_{a_i}}$ by running
  \textsf{RSA-Sig.Sign($\sk_\mathit{CA}$, $a_i || \vk_U$)}, where $||$
  denotes concatenation. &

  \textsf{U.SignM($\sk_U$, $M$):} Given the user's signing key $\sk_U$
  and the message $M$, it runs \textsf{RSA-Sig.Sign($\sk_U$, $M$)} and
  returns the signature $\sigma_M = \mathcal{M}(D)^ {d_U}$.
  \vspace{1em}

  \textsf{OCP.VerifyM($\vk_U$, $\sigma_M$, $M$):} Given the user's
  verification key $\vk_U$, the signature $\sigma_M$ and the message
  $M$, it runs \textsf{RSA-Sig.Verify($\vk_U$, $\sigma_M$, $M$)} and
  returns either \accept \space or \reject.\ \\
 
  \textsf{U.PackA($\vk_\mathit{CA}$, $\sigma_{U_{A^*}}$):} Given the
  CA's verification key $\vk_\mathit{CA}$ and the list of signed
  attributes $\sigma_{U_{{A^*}}} \subseteq \sigma_{U_A}$, it returns
  the packed signature $P_{\sigma_{U_{A^*}}}$. It calculates:
 
  $P_{\sigma_{U_{A^*}}} = \prod\limits_{i=1}^l \sigma_{U_{a_i}} \; mod
  \; N_\mathit{CA}$
 
  where $\sigma_{U_{a_i}} \in \sigma_{U_{A^*}}$ and $l =
  |\sigma_{U_{A^*}}|$. &
 
  \textsf{OCP.VerifyA($\vk_\mathit{CA}$, $\vk_U$,
    $P_{\sigma_{U_{A^*}}}$, $A^*$):} Given the CA's verification key
  $\vk_\mathit{CA}$, the user's verification key $\vk_U$, the packed
  signature $P_{\sigma_{U_{A^*}}}$ and the list of attributes $A^*
  \subseteq A$, it returns either \accept \space or \reject. It
  checks if

  $\prod\limits_{i=1}^l \mathcal{H}(a_i || \vk_U) \stackrel{?}{\equiv}
  (P_{\sigma_{U_{A^*}}})^{e_\mathit{CA}} \; mod \; N_\mathit{CA}$

  where $a_i \in A^*$ and $l = |A^*|$. \\

\hline
 
\end{tabularx}
\caption{\label{fig:own:scheme} Details on the algorithms of the data
  ownership scheme.}

\vspace{-2mm}

\end{figure*}

\begin{definition}[Data ownership scheme]
\label{defAlgorithm}
The data ownership scheme \textsf{DATA-OWN} is \emph{a tuple of
  algorithms} \textsf{$\langle$Init, KeyGen, CA.SignA, U.SignM,
  OCP.VerifyM, U.PackA, OCP.VerifyA$\rangle$}. The definition of the
  algorithms can be found in Figure~\ref{fig:own:scheme}.
\end{definition}

\begin{lemma}[Correctness] 
  Informally speaking, \textsf{OCP.VerifyA($\vk_\mathit{CA}$, $\vk_U$,
    $P_{\sigma_{U_{A^*}}}$, $A^*$)} will always return \accept \space
  if the list of signed attributes that are packed by the user are the
  same as the list of attributes $A^*$ provided by the user to the
  OCP. More formally,
\begin{center}
  \textbf{Pr}[\textsf{OCP.VerifyA($\vk_\mathit{CA}$, $\vk_U$,
    $P_{\sigma_{U_{A^*}}}$, $A^*$)} = \accept] = 1
\end{center}
\end{lemma}

The claim easily follows from the homomorphic property of exponentiation
modulo $N$. We analyze the security properties of \framework in Appendix~\ref{sec:security}.



\section{Performance Analysis}
\label{sec:performance}
In this section, we provide implementation details for all components
that we newly developed for \framework and name libraries that this
implementation relies on. We subsequently evaluate the performance overhead of this implementation for each involved component (CA,
user client, and OCP).

\subsection{Implementation Details and Evaluation Parameters}

\paragraph{Components of the Implementation}
The implementation prototype is written in Java. To reflect the
different involved participants, the implementation consists of three
components: a module for the CA (CA-module), a module for the OCP
(OCP-module) and a module for the user client (user-module). For the
sake of simplicity, the prototypical implementation assumes that the
OCP and the indexing system are managed by the same organization; this
avoids an additional trust level between these institutions and allows
us to concentrate on the performance measurements. The size of each of
these modules (without included libraries; see below) is below 5 KB.

\paragraph{Libraries Used}
Our prototypical implementation relies on several existing open source
libraries. First, we include the Stanford NER
library~\cite{NERlibrary} for identifying personal information in the
textual article. The NER library is of size 3.2 MB and the NER
classifier, for covering seven distinct classes of data, requires 16.6
MB. Second, we rely on OpenCV (Open Source Computer Vision Library),
an open source computer vision and machine learning
library~\cite{opencv}, for face detection and recognition. Finally, we
include the AIDA (Accurate Online Disambiguation of Named Entities)
framework~\cite{AIDA} to achieve ownership disambiguation. In our
experiments, we used the AIDA framework itself and its corresponding
web service, which works with entities registered in the
DBpedia~\cite{DBpedia} or YAGO~\cite{YAGO} knowledge base.

\paragraph{Evaluation Parameters}
We have evaluated the performance of the implementation on a dataset
of 150 news articles that we randomly crawled from the international
news agency Reuters\footnote{\url{http://www.reuters.com/}}, using the
Java-based web crawler \emph{crawler4j}~\cite{crawler4j}. These
articles cover different topics and range from 1 K to about 10 K
words; the average length is 1.9 K words per article. The actual
experiments were run on a standard notebook with 2.5 GHz dual-core
processor and 8 GB RAM. The experimental results described below
constitute the average over 100 independent executions. Network
latency was not considered in the experiments.

\subsection{Evaluating the CA-Module}
Evaluating the performance of the CA-module consists of measuring the
overhead of attribute certification.

\begin{wrapfigure}{r}{0.54\textwidth}
	\vspace{-8mm}
	\begin{center}
		\includegraphics[width=.45\textwidth]{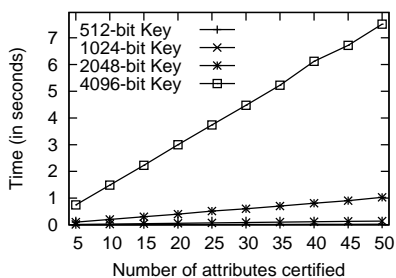}
	\end{center}
	\vspace{-2mm}
	\caption{Evaluation of the CA-module: Performance overhead for certifying user attributes.}
	\vspace{-5mm}
	\label{fig:attribute-gen}
\end{wrapfigure}

\paragraph{Attribute Certification}
Figure~\ref{fig:attribute-gen} illustrates the computational overhead
for certifying user attributes. In our experiment, we generated up to
50 attributes and considered CA's signing keys of varying size,
ranging from 512 to 4096 bits. As we expected, certification time
grows linearly in the number of attributes. For the most complex cases
under consideration --- the CA signing 50 attributes, and thus far more
than what a user would typically maintain, using a signing key of size
4096 bits --- the attribute certification took 7.5 seconds.  For
smaller numbers of attributes, or for all smaller key sizes, this
certification takes less than a second. Since attributes are
typically certified only once per user, this computational overhead
should be acceptable as a one-time upfront effort.

\subsection{Evaluating the User-Module}
Evaluating the user-module is performed in two steps: identifying
suitable attributes in the given sample texts, and pre-processing these
attributes for the subsequent ownership-proof phase.

\begin{figure*}[htp!]
\centering
\subfigure[NER overhead]{
\includegraphics[width=.45\textwidth]{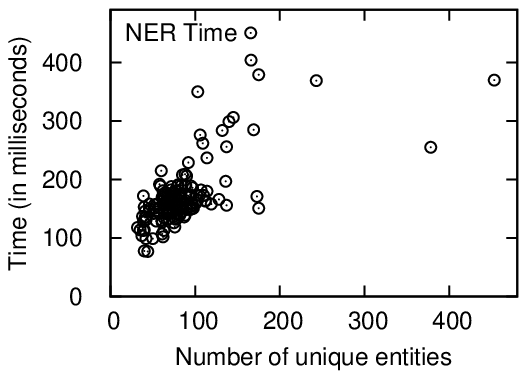}
\label{fig:ner}
}
\subfigure[Packing overhead]{
\includegraphics[width=.42\textwidth]{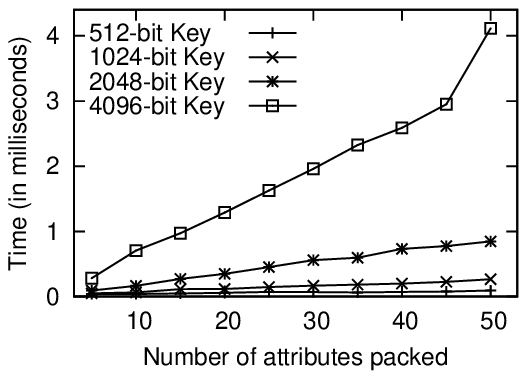}
\label{fig:attribute-pack}
} \vspace{-1em}\caption{\label{fig:user}Evaluation of the user-module:
  Performance overhead of \subref{fig:ner} identifying personal
  information and \subref{fig:attribute-pack} for packing user
  attributes.}
\end{figure*}

\paragraph{Identifying Attributes}
As explained in Section~\ref{sec:approach}, the user-module
pre-processes the article using NER techniques and appropriately
selects all entities that are necessary for the identification
process. We evaluate the performance of the user-module on the
aforementioned 150 news articles from Reuters, and measure the time
required to identify and extract all entities. The results are
depicted in Figure~\ref{fig:ner}. The performance overhead varies from
77 to 814 millisecond (ms), with an average of 174 ms per article. The
number of unique entities in the articles ranges from 43 to 590,
where the average number of unique entities per article is 135.

\paragraph{Attribute Packing}
After identifying all personal attributes in a given news article, the
user-module pre-processes a set of signed attributes as required for
the ownership proof. This pre-processing in particular reduces the
number of exponentiations that are required to verify the attributes
for the OCP, and thereby avoids a potential bottleneck. In the
performance measurement, we again considered up to 50 attributes and
varying key sizes. As shown in Figure~\ref{fig:attribute-pack}, the
time for this pre-processing increases linearly in the number of
attributes, with an additional overhead for larger key sizes. For the
maximum of 50 attributes, the pre-processing only took between 0.1 ms
(for a 512-bit key) and 4.1 ms (for a 4096-bit key).

\paragraph{Message Signing}
The user client signs the message using her signing key. For this
experiment, we considered the aforementioned 150 news articles. Consider the overhead of
signing a message with a signing key of size 1024 bits. Depending on
the size of the article, the signing took between 2.8 and 3.8 ms, with an
average of 2.9 ms per article.

\subsection{Evaluating the OCP-Module}
We split the performance evaluation of the OCP-module into two parts:
First, we evaluate the time required to verify the validity of
requests for varying parameters: for varying numbers of articles, for
varying number of attributes, and for varying verification requests.
Second, we evaluate the time required to decide whether the request is
legitimate, \ie whether the document under consideration affects the
user's data, either by means of entity disambiguation or face
recognition.

\begin{figure*}[ht!]
	\centering
	\subfigure[Overhead of message verification]{
		\includegraphics[width=.4\textwidth]{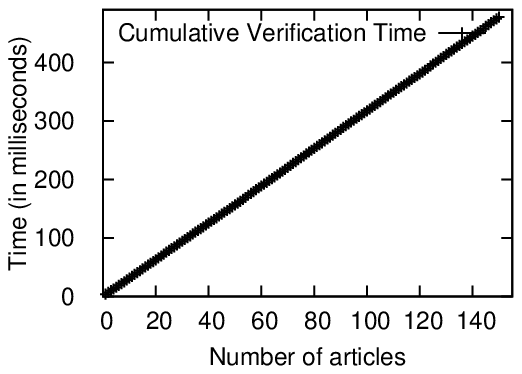}
		\label{fig:article-verify}
	}
	\subfigure[Overhead of attribute verification]{
		\includegraphics[width=.37\textwidth]{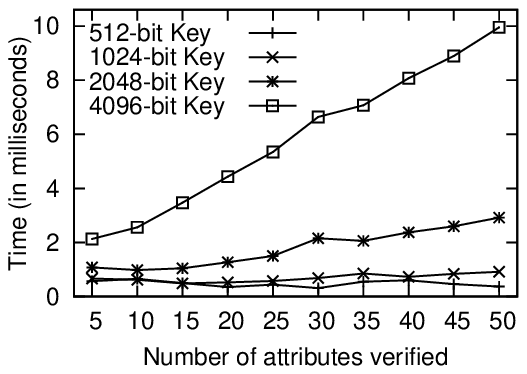}
		\label{fig:attribute-verify}
	}
	\subfigure[Overhead of request verification]{
		\includegraphics[width=.37\textwidth]{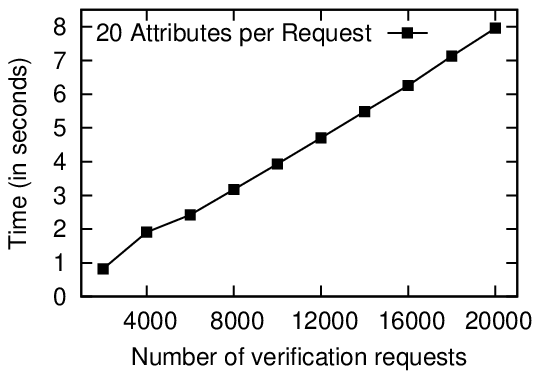}
		\label{fig:verification-requests}
	}
	\subfigure[Overhead of entity disambiguation]{
		\includegraphics[width=.4\textwidth]{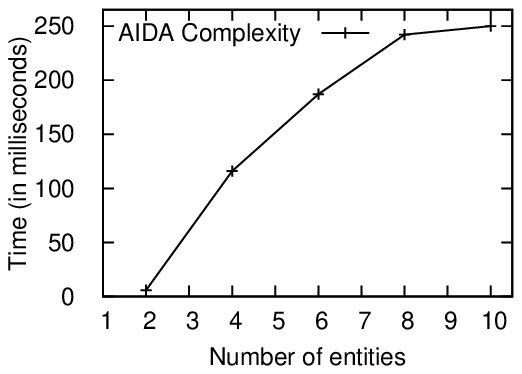}
		\label{fig:aida}
	} \vspace{-0em}\caption{Evaluation of the OCP-module: Performance
	overhead of \subref{fig:article-verify} verifying the messages,
	\subref{fig:attribute-verify} verifying user attributes signed by the CA, \subref{fig:verification-requests} verifying user requests and \subref{fig:aida} running entity disambiguation.}
\label{fig:ocp}
\end{figure*}

\paragraph{Validating the User Request}
Upon receiving a signed message from a user, the OCP verifies the
validity of the signature using the user's verification key. This
verification time (with a 1024-bit key) ranges from 2.9 to 4.3 ms with
an average of 3.2 ms per article. Figure~\ref{fig:article-verify}
illustrates the cumulative verification time to verify up to 150
articles. It grows linearly, so verifying message validity for 150
articles takes the OCP less than 0.72 seconds.

Similarly, Figure~\ref{fig:attribute-verify} displays the time
required to verify a certain number of signed user attributes. Recall
that the user sends a packed version of her signed attributes to ease
the verification task of the OCP. Still, the OCP needs to calculate
the hash of each individual attribute and multiply all hashes together
before being able to verify the signature based on the packed version.
Verifying 50 user attributes takes 0.37 ms (for a 512-bit key)
and 10 ms (for a 4096-bit key), respectively. 
For $l$ attributes, the packed version is at least $l-2$ exponentiations faster than verifying each attribute individually.

Finally, Figure~\ref{fig:verification-requests} shows the
performance overhead for verifying a certain number of user
requests. In our experimentation, we assumed that every request
requires the verification of 20 attributes, each one signed with a key
of size 1024 bits. To measure the performance overhead, we gradually
increased the number of user requests from 2000 to 20,000 and observed
an (essentially linearly-growing) overhead from 0.824 to 7.96 seconds. 
Processing a single verification request with $20$ attributes took less than 0.4 ms on average.

The overall computational overhead of the OCP-module is a combination of the message verification and the attribute request verification, each one incurring on average 3.2 ms and 0.4 ms, respectively. Therefore, our implementation manages to process a
removal request within 3.6 ms. In summary, it allows the OCP to handle
278 requests per second (using the standard laptop that we based
these experiments on).

\paragraph{Eligibility of the User Request}
Identifying whether the requested article indeed contains personal
data of the requesting user relies on appropriate entity
disambiguation. Figure~\ref{fig:aida} illustrates the performance
overhead for entity disambiguation with up to 10 entities.

Recall that we require the user client to run the face detection algorithm
and select the appropriate face and send
it the OCP along with the standard request. The performance overhead
of the face recognition algorithm depends on multiple factors such as
the picture resolution and the face position in the picture. In our
experiments, we have chosen pictures with well-defined frontal
faces. The resolution of the pictures is up to 3072 x 4608 pixels with
an average size of 4 MB. Having all these predefined conditions, the
runtime of the face recognition algorithm stays in the range of 150 to
300 ms.

The overall performance overhead, comprising both entity
disambiguation and image recognition, currently constitutes the
bottleneck for verifying the validity of removal requests in the
OCP-module. Currently, we are exploring further optimization here.



\section{Discussion}
\label{sec:discussion}

\paragraph{Deployability and Usability}
In order to deploy our solution, \framework requires a national or
local government-wide CA that issues credentials to citizens. We argue
that this requirement does not limit practicality of our approach
because the issuance of such credentials is already part of an EU
standard~\cite{EuSignatures}, implemented by some member states and
meant to be adopted by all the EU member states~\cite{GermanID,
  EuID}. The European EID standards also enable the use of digital
credentials for Internet communication (\eg for online
shopping)~\cite{GermanID} which also strengthens usability for
\framework's developers as well as end-users.

\paragraph{Scope of Eligibility}
First, it is a hard problem to decide on the eligibility of an
ownership claim if two persons have the same attributes, e.g.,
name. \framework addresses this issue by using attributes that in
combination should be sufficiently unique for most people. Second, our
framework cannot decide whether a piece of content is of public
interest (such information falls into the category of freedom of the
press) and outweighs the privacy interest of an individual. This
decision is a legal assessment. This is outside of the scope of
\framework and subject to ongoing research about the automation of
legal assessments~\cite{BaBeHoMa:2015}. 


\paragraph{Privacy and Availability}
The OCP could be a third-party service or managed by the search engine
provider. From a privacy point of view, the latter setup may reveal
personal information about citizens. However, we argue that a search
engine provider does not learn more than what is already available in
the article. This is because \framework follows a principle of least
privilege, where only those particular attributes that are present in
the article are sent to the OCP. The collection of information and
verification makes the OCP a key component of \framework. The
availability of the OCP becomes essential in the long-run success of
\framework. Therefore, to prevent a single point of failure, we can
consider deploying multiple instances of the OCP.

\paragraph{Robustness}
\framework relies on NLP and image recognition techniques. The NLP
technique we use in our framework is simple and sufficiently robust in
practice. Concerning robustness of the image recognition technique,
recent research has shown that automated face recognition is almost
comparable to human face recognition
accuracy~\cite{DeepFace}. Therefore, when the removal request includes
a picture that uniquely identifies the user with a certain confidence
(part of the deployed policy), our framework can easily approve the
removal request.

\section{Conclusion}
\label{sec:conclusion}
In this work, we have introduced a universal framework, called \framework, providing the foundation to support the enforcement of the \emph{right to be forgotten} in a scalable and automated manner both for users and indexing systems. The framework enables a user to automatically
identify personal information in a given article and the indexing system to automatically verify the user's eligibility. The framework also achieves censorship-resistance, \ie
users cannot blacklist a piece of data unless it affects them personally. This is accomplished using the government-issued digital
credentials as well as applying the entity disambiguator technique.
 We have conducted comprehensive evaluations of \framework on existing articles, showing that the framework incurs only minimal
overhead and is capable of handling 278 removal requests per second on a standard notebook (2.5 GHz dual core). 
In these evaluations, we have observed that the remaining performance
bottleneck on the OCP is caused by the entity disambiguator (\ie AIDA)
and the face recognition (\ie OpenCV) algorithms. We believe that
optimized versions of both could help in significantly improving the
performance.

For future work, we plan to improve \framework's accuracy and overall coverage for proving affectedness. Following the principle of reCAPTCHA digitizing books \cite{Von:2008:recaptcha}, improving the accuracy of NER by taking into account the user client tagging constitutes a promising approach.
Another promising research direction is to analyze the assessment of lawfulness and
automate the application of future guidelines for the \emph{right to be forgotten}. Staying close to the precedent, this would also require semantically analyzing the article to determine if its content violates privacy rights, \eg by being outdated or by containing sensitive information for the entity requesting removal.


{\footnotesize \bibliographystyle{splncs03}
\bibliography{bib/myBib}}

\appendix

\section{Security Analysis}\label{sec:security}
The framework is supposed to achieve three security objectives:
minimal disclosure, request unforgeability and censorship-resistance
(cf. Section~\ref{sec:goals}). In the following, we show that we
achieved these goals.

\paragraph{Minimal disclosure} 
Minimal disclosure in this context is
the minimization of knowledge increase for the indexing system in
order to verify eligibility of a request. In the case that OCP and IS
are separate, this is given since the IS only receives a token from
the OCP through the user. This token does not need to contain any
information about the user. However, if the OCP and the IS collide, it
has to receive the input to run \textsf{OCP.VerifyA}. 

A user who wants to hide her verification key or credentials could
potentially use interactive zero-knowledge proofs on top of our
construction. This, however, would sacrifice efficiency and would not
improve the disclosure of information. The reason is that the
verification key is basically a pseudonym, \ie it is only linked to
the attributes that we send. Thus, the only need to minimize is the
sending of attributes. In \framework, we only send those attributes that are
indeed necessary for proving that the user is affected, \ie that already occur in
the link we report. We implement this by sending only
subsets.\footnote{We stress again that we only show affectedness of the user. Arguing
about the legal implications and whether this
  minimization of data is sufficient in order to apply them is beyond the scope
of this (and all existing) work.}

\paragraph{Request Unforgeability} For unforgeability we show that
even the user cannot construct a message that verifies without having
a signature on every single attribute. As a consequence, the user
cannot show that she is affected by content concerning other users'
attributes.

\begin{theorem}[Request Unforgeability] If \textsf{OCP.VerifyA}
  returns $\accept$ then the packed attributes correspond to the set
  $A^*$. More formally, every $\mathcal{A}$ that has access to a
  signing oracle $\mathcal{S}$ with public key $\vk$ can only generate
  $P$ for subsets $A^*$ of all signatures $A$ requested from
  $\mathcal{S}$.
\end{theorem}

\paragraph{Proof} 
Let $A$ be the set of queried attributed signatures of the adversary
$\mathcal{A}$ for a given execution. Assume there is a set
$(B,P)\ot\mathcal{A}^\mathcal{S}(\vk_u)$ such that
\textsf{OCP.VerifyA($\vk$, $\vk_U$, $P$, $B$)} = $\accept$  and
$B\not\subseteq A$. So there exists  $b^* \in B$ such that $b^*\not\in A$.  
Then there also exists an adversary $\mathcal{A^*}$
that queries $A\cup B\backslash\{b^*\}$, \ie $b^*$ is the only
unqueried attributed in $B$. Since \textsf{OCP.VerifyA($\vk$, $\vk_U$,
  $P$, $B$)} = \accept, it follows that $P^{e_\mathit{CA}}\equiv
\prod_{b\in B} \mathcal{H}(b || \vk_U)$ by construction. Since we
queried all $b$ except $b^*$ in $B$, we can compute $\sigma :=
P/\prod_{b\neq b^*\in B} \mathcal{H}(b_i ||
\vk_U)^{d_\mathit{CA}}$. For this $\sigma$, we have
$\sigma^{e_\mathit{CA}} = \mathcal{H}(b^* || \vk_U)$. However, this
contradicts the Chosen Message Attack (CMA) security of the underlying signature scheme. Thus,
the adversary $\mathcal{A}$ cannot exist. $\blacksquare$

\paragraph{Censorship-Resistance}
Finally, we have to ensure that the overall system does not enable any
user to censor, \ie to successfully report data that she is not affected by. There are two possible approaches. First, we could do a reduction
proof to the CMA-security of the signature scheme as done for the
request unforgeability.\footnote{Such a proof would look like this:
  Assume censorship is possible. That means there is an execution that
  ends with a successful report at the indexing system without the
  user reporting the data. Therefore, there was a $D_U$ sent to the
  IS that verifies with the key of the OCP. Either the OCP signed
  $D_U$ or there is a contradiction to the signature scheme's
  CMA-property. Consequently, the OCP signed $D_U$ and since we assume
  the OCP to be trustworthy, it means that the OCP received an $M,
  \sigma_M$ from a user and verified it. Here, either the user's
  signature $\sigma_M$ was forged (contradicting the CMA-property of
  the signature scheme) or the user forged a message $M$ that verifies
  (contradicting the request unforgeability proven before). It follows
  that the user could not have generated such a request, proving
  censorship-resistance.\\ While this argumentation sounds plausible,
  it does not consider every possible interleaving or repetition of
  executions. In contrast, tool support offers a trustworthy guarantee that we did not overlook any execution generated by these processes.}  Second, we can
formulate the protocol in the applied $\pi$-calculus and automatically
verify the properties of interest using tool support. The outcome can
then be leveraged from the protocol to the implementation by using
computational soundness which links symbolic execution traces to
computational execution traces. Thus, we can use tools for symbolic
verification and the outcome transfers to the implementation. In what
follows, we pursue the second approach since the protocol is easy to express
and verify using state-of-the-art verification tools.
\begin{sloppypar}
The applied $\pi$-calculus defines a way of modeling processes
(\verb+P,Q+). Thereby, the calculus gives constructs for parallel
execution of processes (\verb+P|Q+), for repetition of processes
(\verb+!P+), for communication between processes (\verb+in(chan,msg),out(chan,x)+) and for restricted computation. The
restriction is that only symbolic constructors
(\verb+let x=sig(sk,msg)+) and destructors
(\verb+let m=verify(vk,sig)+) can be used to modify terms which
consist of symbols. The difference is that constructors create
symbolically larger symbols, \ie in the example $x$ will be handled
as the symbol \verb+sig(sk,msg)+ whereas destructors can give
reduction rules to remove or replace constructors. Finally, for
symbols, there are two classes, publicly known symbols and freshly
introduced symbols (\verb+new N; P+) which are unequal to all other
symbols.
\end{sloppypar}
For the sake of exposition, we briefly describe the process for the indexing
system. The system receives a message and verifies it with the
corresponding key (computational soundness requires
the key to be part of the signature). We then verify the first part of
the signed message with the verification key of the OCP. This message
must be the user's verification key and the requested data, \ie we
check for equality before the IS is convinced that the signer is
affected.
\begin{verbatim}
let IS = in(ch,x); let tmpkey = vkof(x);
   let (c,reportData) = verifySig(tmpKey, x) in 
   let (sigKey, sigData) = verifySig(vkOCP, c) in
   if reqData==sigData then if tmpKey==sigKey then
   event affected(sigKey,reqData).
\end{verbatim}\vspace{-4em}\hfill \vspace{1em}

The end of the process is a so-called event. These events have no semantic
meaning in the calculus, but can be used by the model checker to prove
certain properties of the protocol. In our example, the event
symbolizes the belief of the IS that the user with verification key
\verb+sigKey+ is affected by the data \verb+reqData+. The model
checker answers queries such as \verb+query ev:affected(k,d).+ which formalizes that the model checker can prove that this event can be reached in the
protocol execution.

Censorship-resistance can be formulated as a sequence of events that
has to occur whenever the IS thinks a user is affected, \ie whenever
a request is considered to affect the requesting user, the OCP has
verified that this data belongs to the user that sends the
request. This can be done by two queries of the form
\verb+query ev:affected(key,d) ==> ev:VerifiedOw(x,key,d).+ meaning
that whenever the event \verb+affected+ occurs there has to be a
corresponding event verifying the ownership beforehand. Analogously,
we prove that the ownership verification is preceded by the attribute
verification of the CA.

The complete formalization in the applied $\pi$-calculus can be found online at the project website\footnote{\url{https://infsec.cs.uni-saarland.de/projects/oblivion/}}. 
The protocol verification takes 8 ms.

\paragraph{Other Security Goals}
In order to prevent a replay attack, the user includes the timestamp in
her request. One can argue that a replay attack is not an issue because
it is a legitimate request by the authorized user. However, we
consider that a replay attack could harm the credibility of the user if an
adversary launches it to mount a Denial-of-Service (DoS) attack on the OCP.


\end{document}